\documentclass[12pt]{article}
\usepackage[dvips]{graphicx}
\usepackage{latexsym}
\usepackage[abs]{overpic}
\usepackage{wrapfig}
\usepackage{amssymb}
\usepackage{amsmath}
\usepackage{color}
\usepackage{bm}

\input epsf
\newcommand{\skyp}[1]{}

\def\rem#1{}
\def\ep{{\epsilon}}

\def\mS{\cS}

\def\sig{{\sigma}}

\def\1{{I}}
\def\2{{I\hspace{-.1em}I}}

\def\rem#1{}

\newcommand\non{\nonumber \\}
\newcommand{\bel}{\begin{eqnarray}}
\newcommand{\ee}{\end{eqnarray}}

\makeatletter
\@addtoreset{equation}{section}
\makeatother

\newcommand{\Gr}{\mathrm{Gr}}

\newcommand{\cO}{\mathcal{O}}
\newcommand{\cQ}{\mathcal{Q}}
\newcommand{\cS}{\mathcal{S}}
\newcommand{\bC}{\mathbb{C}}
\newcommand{\Xv}{\check{X}}
\newcommand{\tst}{$\mathbf{27}^3$} % twenty-seven to the third
\newcommand{\tsst}{$\mathbf{27}^{*3}$} % twenty-seven star to the third

\begin{document}

\begin{titlepage}

\bigskip
\hfill\vbox{\baselineskip12pt
\hbox{KIAS-P16013}
%\hbox{}
}
\bigskip\bigskip\bigskip\bigskip\bigskip\bigskip\bigskip\bigskip

\begin{center}
\Large{ \bf
Equivariant A-twisted GLSM and Gromov--Witten invariants
of CY 3-folds in Grassmannians  
}
\end{center}
\bigskip
\bigskip

\bigskip
\bigskip
\centerline{ \large  Kazushi Ueda$^{1}$ and Yutaka Yoshida$^{2}$}
\bigskip
\medskip
\centerline{$^{1}$ \it Graduate School of Mathematical Sciences,
The University of Tokyo,}
\centerline{\it  3-8-1 Komaba
Meguro-ku
Tokyo
153-8914
Japan. }
\centerline{kazushiATms.u-tokyo.ac.jp }
\bigskip
\bigskip
\centerline{$^{2}$  \it School of Physics, Korea Institute for Advanced Study (KIAS),}
\centerline{\it  85 Hoegiro Dongdaemun-gu, Seoul 02455, Republic of Korea. }
\centerline{ yyoshidaATkias.re.kr}
\bigskip
\bigskip

\begin{abstract}
We compute genus-zero Gromov--Witten invariants
of Calabi--Yau complete intersection 3-folds in Grassmannians
using supersymmetric localization
in A-twisted non-Abelian gauged linear sigma models.
We also discuss a Seiberg-like duality
interchanging $\Gr(n,m)$ and $\Gr(m-n,m)$.
\end{abstract}
\end{titlepage}

\newpage
\baselineskip=18pt

\tableofcontents

\section{Introduction}

In a Calabi--Yau (CY) compactifications of $E_8 \times E_8$ heterotic string theory
with the standard embedding,
there are four types of the Yukawa couplings
$
 (\mathbf{1}^3, \mathbf{1} \cdot \mathbf{27} \cdot \mathbf{27}^*,
   {\mathbf{27}}^{* 3}, {\mathbf{27}}^{ 3})
$
in the 4-dimensional low energy effective theory.
It is known that \tst-type Yukawa couplings
do not receive either  loop or  world-sheet instanton corrections.
On the other hand,
the \tsst-type Yukawa couplings receive  corrections
coming from world-sheet instantons.
It was conjectured in \cite{Candelas:1990rm} that
the world-sheet instanton corrected \tsst-type Yukawa couplings
can be explicitly computed
from the \tst-type Yukawa couplings
of the mirror manifold.

A generalization of A-twisted gauged linear sigma models (GLSMs)
with one omega-background parameter on a 2-sphere $S^2$
has been constructed in \cite{Closset:2014pda}.
Recently, the supersymmetric localization computations on this geometry have been performed in \cite{Closset:2015rna} (See also \cite{Benini:2015noa}).
It gives an explicit formula
for cubic correlation functions of scalars in the vector multiplet,
which conjecturally give the Yukawa couplings of the mirror
when the omega-background parameter is set to zero.
An interesting point here is that
one can compute the \tst-type Yukawa couplings
without knowing the mirror manifold.
This formula goes back to \cite{Morrison:1994fr}
when the gauge group is Abelian.
A mathematical conjecture in the Abelian case,
called \textit{toric residue mirror conjecture},
is formulated in \cite{BM1,MR2019144}
and proved in \cite{Szenes, Borisov, MR2147350}.
The formula obtained in \cite{ Closset:2015rna}
works also for non-Abelian gauge theories, and
can be regarded as a generalization of toric residue mirror conjecture
to CY manifolds in non-toric manifolds.
We also give an explicit computation of the mirror map
in a framework of A-twisted GLSMs with omega-background parameter.
This allows us to give a conjectural computation
of the genus-zero Gromov--Witten invariants
of the CY manifolds.

CY 3-folds with one-dimensional K\"ahler moduli
defined as complete intersections 
of zero loci of sections of equivariant vector bundles
on Grassmannians
are classified in \cite{Inoue1}.
In this paper,
we realize some of them
as phases of GLSMs,
and compute the Yukawa couplings
in terms of A-twisted GLSM.
This allows us to give a conjectural computation
of genus-zero Gromov--Witten invariants
of such CY 3-folds.
The result agrees with a mathematically rigorous treatment
obtained earlier in \cite{Inoue2}
based on Abelian/non-Abelian correspondence
\cite{MR2110629,MR2369087,MR2367022}.

This paper is organized as follows:
In Section 2,
we briefly review GLSMs with unitary gauge groups and
their relation to CY 3-folds in Grassmannians.
In Section 3,
we discuss the mirror map
for  CY 3-folds in  Grassmannians
in terms of A-twisted GLSMs with omega background.
In Section 4,
we use the method in Section 3
to compute the genus-zero Gromov--Witten invariants
of some CY 3-folds in Grassmannians. 
In Section 5,
we discuss a Seiberg-like dual description
of the Yukawa coupling.
The last section is devoted to a summary.

%%%%%%%%%%%%%%%%%%%%%%%%%%%%%%%%%%%%%%%%%%%%%%%%%%%%%%%%%%%%%%%%%%
%%%%%%%%%%%%%%%%%%%%%%%%%%%%%%%%%%%%%%%%%%%%%%%%%%%%%%%%%%%%%%%%%%
%%%%%%%%%%%%%%%%%%%%%%%%%%%%%%%%%%%%%%%%%%%%%%%%%%%%%%%%%%%%%%%%%%
\section{GLSMs and CY 3-folds in  Grassmannians}
\label{section2}

In this section,
we consider 2d $\mathcal{N}=(2,2)$ GLSMs
\cite{Witten:1993yc}
with gauge group $G=U(n)$
which flow to infrared non-linear sigma models (NLSMs)
with large positive Fayet--Iliopoulos parameter (FI-parameter) $\xi \gg 0$.
The target spaces are given by the Higgs branch moduli of GLSMs.
In this paper,
we study the case where the target spaces are CY 3-folds
defined as complete intersections of zeros of sections of vector bundles
constructed from the dual of the universal subbundle $\cS$
on Grassmannians. 

The matter multiplets consist of  $m$ fundamental chiral multiplets
$\Phi^i$ for $i=1,\cdots,m$
and chiral multiplets $P_l$ for $l=1,\cdots s$
in the gauge representation $R_l$.
When all the $P_l$ are absent,
the D-term  vacuum condition   for $\Phi^i$ in the Higgs branch
defines the Grassmannian $\Gr(n,m)$
in the positive FI-parameter region.
The introduction of $P_l$ modifies the D-term vacuum condition
to the total space of  the vector bundle
on the Grassmannian  $\oplus_{l=1}^s E_{R_l} \to \Gr (n,m)$
with appropriate choices of gauge representations $R_l$.
The vector bundle $E_{R_l}$ is determined by the gauge representation $R_l$.
For examples,  relations between $R_l$ and $E_l$ are
\bel
&&R_l={\mathbf{n}}^* \longleftrightarrow  E_l =\mS, \\
&&R_{l}=\mathrm{det}^{-q} \longleftrightarrow E_l =\cO(-q)=\cO(-1)^{\otimes q} , \\
&&R_l=\mathrm{Sym}^q {\mathbf{n}}^* \longleftrightarrow E_l =\mathrm{Sym}^q {\mS}, \\
&&R_l=\Lambda^q {\mathbf{n}}^* \longleftrightarrow E_l =\Lambda^q {\mS}, \\
&& \cdots \nonumber
\ee
Here ${\mathbf{n}}^*$,
$\mathrm{det}^{-1}$,
$\mathrm{Sym}^q {\mathbf{n}}^*$, and
$\Lambda^q {\mathbf{n}}^*$ represent
the anti-fundamental representation,
the inverse of the determinant representation,
the $q$-th symmetric products of the anti-fundamental representation, and
the $q$-th anti-symmetric products of the anti-fundamental representation
of the gauge group $U(n)$, respectively.
The bundle $\mS$ is the universal subbundle and
$\cO(-1)=\Lambda^n \cS$ is the inverse of the determinant line bundle. 

We introduce the following superpotential term 
\bel
W(P,\Phi)=\sum_{l=1}^s P_l G_l (\Phi).
\label{superpt}
\ee
Here $G_l(\Phi)$ is a homogeneous polynomial in $\Phi^i$
which belongs to the complex conjugate representation of $R_l$.
The polynomial $G_l(\phi)$ defines a section of the bundle $E^{*}_{l}$ on $\Gr(n,m)$,
where $E^*_l$ is the dual bundle of $E_l$.
The F-term equations of  (\ref{superpt}) are given by 
\bel
&&G_{l} (\phi)=0, 
\label{eq:Fterm1} \\
&&\sum_{l=1}^s p_l \frac{\partial G_l}{ \partial \phi_i}=0. 
\label{eq:Fterm2}
\ee
If the equation \eqref{eq:Fterm1} defines a smooth complete intersection
in the Grassmannian,
then it follows from the Jacobian criterion for smoothness
that the rank of the matrix $\left( \frac{\partial G_l}{ \partial \phi_i} \right)$
appearing in \eqref{eq:Fterm2}
is equal to the sum $\sum_{l=1}^s \dim R_l$ of the dimensions of $p_l$,
so that the only solution to \eqref{eq:Fterm2} is $p_l=0$ for all $l$.
The F-term and D-term equations reduce to
\bel
&&\sum_{i=1}^m \phi^i (\phi^i)^{\dagger}= \xi \mathbf{1}_n, \quad G_{l} (\phi)=0.
\ee
Then the GLSM flows to non-linear sigma model whose target space is given by 
the complete intersection of zero section of $E^{*}_l$ in the Grassmannian $\Gr(n,m)$;
\bel
X^{n,m}_{\oplus^s_{l=1} E^{*}_{l}} :=\{ [\phi^i] \in \Gr(n,m) | G_l(\phi)=0, ~ l=1,2, \cdots, s \}.
\label{targetsp}
\ee
The complex dimension of $X^{n,m}_{\oplus^s_{l=1} E_{l}}$ is given by
\bel
\mathrm{dim}_{\bC} X^{n,m}_{\oplus^s_{l=1} E_{l}}
=m n-n^2-\sum_{l=1}^s \dim R_l.
\label{dimensions}
\ee
Recall that the dimensions of the symmetric and anti-symmetric
representations are given by
\bel
 \dim R_l=
\begin{cases}
 \dfrac{(n+q+1)!}{q ! (n-1)!} & R_l =\mathrm{Sym}^q {\mathbf{n}}^*, \\
 \dfrac{n!}{(n-q)! q!} & R_l =\Lambda^q {\mathbf{n}}^*.
\end{cases}
\ee
Since we are interested in the cases where the target spaces are CY manifolds,
the axial anomaly has to be canceled:
\bel
2\pi  i \partial_{\mu} J^{\mu}_A = m \mathrm{Tr}_{\mathbf{n}} F_{12}+\sum_{l=1}^s \mathrm{Tr}_{R_l} F_{12} =0.
\ee
We will give a computation
which conjecturally gives genus-zero Gromov--Witten invariants of CY 3-folds
realized as axial anomaly free non-Abelian GLSMs.

%%%%%%%%%%%%%%%%%%%%%%%%%%%%%%%%%%%%%%%%%%%%%%%%%%%
\section{Equivariant A-twisted GLSM on two sphere and mirror symmetry}

Supersymmetric backgrounds in two dimensions have been studied
from a rigid limit of linearized new minimal supergravity \cite{Closset:2014pda}.
There exists a new supersymmetric background on $S^2$
which is an extension of the topological A-twist
by one omega-background parameter $\hbar$
(equivariant A-twisted GLSM).
If $\hbar$ is set to zero,
the theory reduces to the ordinary A-twisted GLSM on $S^2$.
 
In \cite{Closset:2015rna, Benini:2015noa},
the correlation functions
of gauge invariant operators
coming from the vector multiplet scalar $\sigma$ 
inserted at the north and the south poles of $S^2$
have been evaluated by supersymmetric localization.
For $G=U(n)$, the saddle point value of the $a$-th diagonal component $\sigma$
at the north and the south pole are given by
\bel
&&\sigma_a (x) |_{\mathrm{N}} =\sigma_a-\frac{k_a}{2} \hbar \quad {\rm ( north ~ pole)}, \\
&&\sigma_a (x) |_{\mathrm{S}} =\sigma_a+\frac{k_a}{2} \hbar \quad  {\rm (south ~ pole)}.
\ee
Here $k_a$ for $a=1,\cdots, n$ are the magnetic charges
for the diagonal elements of the gauge fields.
The correlation function of $\langle f(\sigma) |_{\mathrm{N}} g(\sigma) |_{\mathrm{S}}  \rangle$  for $G=U(n)$ is given by
\bel
\langle f(\sigma) |_{\mathrm{N}} g(\sigma) |_{\mathrm{S}}   \rangle 
&&=\frac{1}{n!} \sum_{\mathbf{k} \in \mathbb{Z}^n} z^{\sum_{a=1}^n k_a} \sum_{\sigma_{*}}
%\nonumber \\ && \times 
{\rm JK \mathchar `-Res} (\mathbf{Q}(\sigma_*), \eta) 
Z^{\text{vec}}_{\mathbf{k}} \left(\prod Z^{\text{chiral}}_{\mathbf{k}} \right)
 f \left(\sigma-k \frac{\hbar}{2} \right) g\left(\sigma+k \frac{\hbar}{2} \right). \non
\ee
Here $f(\sigma) |_{\mathrm{N}}$ and $g(\sigma) |_{\mathrm{S}}$ are
gauge invariant operators
constructed from $\sigma$ and inserted on the north and south pole respectively. 
The variable $z$ is the exponential of the complexified FI-parameter
defined as $z := e^{2\pi \sqrt{-1} (\theta+\frac{\sqrt{-1}}{2\pi}  \xi)}$ with theta angle $\theta$.
The factors $Z^{\text{vec}}_{\mathbf{k}}$ and $\prod Z^{\text{chiral}}_{\mathbf{k}}$
are the contributions from the one-loop determinants of the $U(n)$ vector multiplet
and chiral multiplets with magnetic charge $\mathbf{k}=\mathrm{diag} (k_1, \cdots, k_n)$ respectively,
and have the following forms:
\bel
&&Z^{\text{vec}}_{\mathbf{k}}(\sig, \hbar)=(-1)^{\sum_{a < b} (k_a-k_b)}\prod_{1 \le a\neq b \le n} \left(\sigma_a-\sigma_b+|k_a-k_b|\frac{\hbar}{2} \right), \\
&&Z^{\text{chiral}}_{\mathbf{k}} (\sig, \lambda, \hbar)=
 \prod_{\rho \in \Delta(R)} \prod_{j=-\frac{|\rho(\mathbf{k}) -r+1|-1}{2}}^{\frac{|\rho(\mathbf{k}) -r+1|-1}{2}}
(\rho(\sigma)+\lambda+j \hbar )^{-\text{sign}(\rho(k) -r+1)}.
\ee
Here $\Delta(R)$ is the set of weights of $R$. $\lambda$ is the twisted mass. 
${\rm JK \mathchar `-Res}(\mathbf{Q}(\sigma_*), \eta)$ is the Jeffrey-Kirwan residue operation determined by a charge vector $\eta$ at a singular locus $\sigma_*$. 
The variable $r$ is  an integer R-charge of the lowest component scalar
 in the chiral multiplet.
We assign $r=0$ for the fundamental  chiral multiplets which parametrize the coordinates of the target space in the low energy NLSM. This assignment is compatible with 
 the R-charge assignment in an A-twisted NLSM which is relevant to the  Yukawa coupling computation  \cite{Witten:1991zz}. 
Then, according to \eqref{superpt}, the R-charge of  the lowest component scalar $p^l$ in the chiral multiplet $P^l$ is determined to be $r=2$.

For all the GLSMs that we consider in the next section,
the weights $\rho$ of the chiral multiplet $P^l$ satisfy the condition
$\rho(\mathbf{k}) \le 0 $ for $\mathbf{k} \in \mathbb{Z}^n_{\ge 0}$.
Then with the choice of $\eta=(1, \cdots, 1)$,
the Jeffrey-Kirwan residue is the sum of residues at the poles
coming from the fundamental chiral multiplets, and
has following simple form:
\bel
&&\langle f(\sigma)|_{\mathrm{N}} g(\sigma)|_{\mathrm{S}}  \rangle
=\frac{1}{n!} \sum_{\mathbf{k} \in \mathbb{Z}^n_{\ge 0}} ((-1)^{n-1} z)^{\sum_{a=1}^n k_a}
\oint \prod_{a=1}^n \frac{d \sig_a}{2\pi \sqrt{-1}}    f\left(\sigma-k \frac{\hbar}{2} \right) g\left(\sigma+k \frac{\hbar}{2}\right) 
\non &&~~~~~~~~~~~~~~~~ \times 
\prod_{1 \le a\neq b \le n} \left(\sigma_a-\sigma_b+|k_a-k_b|\frac{\hbar}{2} \right) 
\frac{\prod_{l=1}^s \prod_{\rho \in \Delta(R_l)} \prod_{j=\frac{\rho(\mathbf{k})}{2}}^{\frac{-\rho(\mathbf{k})}{2}} (\rho(\sig) +\lambda_l'+j \hbar)}
{\prod_{i=1}^m \prod_{a=1}^n \prod_{j=-\frac{k_a}{2}}^{\frac{k_a}{2}}(\sigma_a +\lambda_i +j \hbar )}. 
\nonumber \\
\label{omegaS2}
\ee
Here $\lambda_i$ is the twisted mass  of $\Phi^i$ and $\lambda_l'$ is the twisted mass  of $P_l$. 
The contour integrals enclose all  the poles $\sigma_a =-j \hbar -\lambda_i$ with 
$i=1,\cdots, m$, $  j=-\frac{k_a}{2}, \cdots, \frac{k_a}{2} $ and $k_a =0,1,\cdots$.

If we set $\hbar=0$,
equivariant A-twist reduces to ordinary A-twist on $S^2$.
In this case,
$\sigma (x)$ is invariant under the supersymmetric transformation
at any point on $S^2$, and
the saddle point value is simply given by a constant configuration $\sigma_a$.
Then the correlation function of $(\mathrm{Tr}\sigma )^{M}$ with $\lambda_i=\lambda_l'=0$ is given by
\bel
&&\langle (\mathrm{Tr}\sigma )^{M}  \rangle_{\hbar=0} =\frac{(-1)^{\frac{n(n-1)}{2}}}{n!} \sum_{\mathbf{k} \in \mathbb{Z}^n_{\ge 0}}  ((-1)^{n-1}z)^{\sum_{a=1}^n k_a}
\oint_{\sigma=0} \prod_{a=1}^n \frac{d \sig_a}{2\pi \sqrt{-1}}    \left(\sum_{a=1}^n \sigma_a \right)^{M}
\non &&~~~~~~~~~~~~~~~~ \times
\prod_{1 \le a < b \le n} \left(\sigma_a-\sigma_b \right)^2  
\frac{\prod_{l=1}^s \prod_{\rho \in \Delta(R_l)}  \rho(\sig)^{-\rho(\mathbf{k})+1} }
{\prod_{a=1}^n \sigma_a^{m(k_a+1)}}.  
\label{omegazero}
\ee
If the target space of the low energy NLSM is a CY 3-fold,
the expectation values $\langle (\mathrm{Tr} \sigma )^M \rangle_{\hbar=0}$
except for $M=3$ are zero, and
it was conjectured that the expectation value $\langle (\mathrm{Tr}\sigma)^3 \rangle_{\hbar=0}$ 
gives the \tst-type Yukawa coupling
\bel
K_{zzz} := \int_{\Xv} \Omega ({z}) \wedge \left({z} \frac{\partial }{\partial {z}}\right)^3 \Omega ({z})
\label{Yukawa}
\ee
in four dimensions compactified
by the mirror manifold $\Xv$.
Here $\Omega ({z})$ is the holomorphic $(3,0)$-form on $\Xv$ and
${z}$ is the complex structure moduli of $\Xv$.
We comment on the over all sign ambiguity and
sign difference between exponentiated FI-parameter and the complex structure moduli
in (\ref{omegazero}) and (\ref{Yukawa}).   
%%%%%%%%%%%%%%%%%%%%
 %%%%%%%%%%%%%%%%%%%%
Since the first  coefficient of (\ref{Yukawa}) in series expansion of $z$ agrees
with the triple intersection number of $X$ which is positive,
the overall sign of (\ref{omegazero})  is fixed
by requiring the positivity of the triple intersection number. 
In general, the exponentiated FI-parameter $z$ has a different sign
from the complex structure.
In the computation of Gromov--Witten invariants,
this sign ambiguity is fixed by requiring the Gromov--Witten invariant to be positive. 
In the models treated in the next section,
the sign shift $z \to (-1)^{m} {z}$ in the GLSM gives the correct sign
of the mirror Yukawa coupling.

When the gauge group is $U(1)$,
the ambient space of the target space is
the complex projective space $\mathbb{P}^{m-1}$, and
(\ref{Yukawa}) was first proposed by \cite{Morrison:1994fr}. 
A mathematical interpretation of this formula was proposed  by \cite{BM1}.
It has been  shown in \cite{Closset:2015rna} that
(\ref{omegazero}) for GLSMs studied in \cite{Hori:2006dk}
gives correct mirror Yukawa coupling
given in \cite{BCFKS}. 
So we expect that (\ref{Yukawa}) also works
for the non-Abelian GLSMs considered in the next section.
 
In order to extract genus-zero Gromov--Witten invariants
from the mirror Yukawa coupling,
we have to rewrite the mirror Yukawa coupling
in terms of the flat coordinate $t$.
The relation between $z$ and $t$ is given by the \textit{mirror map},
defined as the ratio of two period integrals
$I_0 (z)$ and $I_1(z)$ of $\Xv$;
\bel
t=\frac{I_1(z)}{I_0(z)}.
\label{mirrormap}
\ee 
Here $I_0(z)$ is the fundamental period normalized as $I_0(0)=1$, and  
$I_1(z)$ is the period with a logarithmic monodromy
$I_1(z)=I_0 (z)\log z+\tilde{I}_1(z)$ with $\tilde{I}_1(0)=0$.
By solving (\ref{mirrormap}) recursively and
expressing $z$ as a function of $q:=e^t$,
the Yukawa coupling in the flat coordinate
is written as
\bel
&&K_{ttt}=\frac{K_{zzz} (z(q))}{ I^2_0(z(q))} \left(\frac{q}{z (q)}\frac{dz (q) }{d q} \right)^3 =
n_0+\sum_{d=1}^{\infty} n_d d^3 \frac{q^d}{1-q^d}, \\
&&N_d=\sum_{ k | d}  \frac{n_{\frac{d}{k}}}{k^3}.
\ee
Here $n_d$ for $d =0, 1, 2, \ldots$ are the instanton numbers 
of the target space $X$
of genus 0 and degree $d$. $N_d$  for $d =0, 1, 2, \ldots$ are  genus zero  Gromov--Witten invariants. 
In particular, $n_0 = \int_X H^3$ is the triple intersection number
of the hyperplane class $H$ in $X$.

The period integrals and the mirror map can be extracted
from the equivariant A-twisted GLSM as follows.
Motivated by the factorization property of the physical $S^2$ partition function
\cite{Benini:2012ui, Doroud:2012xw, Bonelli:2013mma},
we rewrite correlation functions as follows.   
After redefinitions of integration variables, the expectation values of $(\mathrm{Tr} \sigma)^M$ inserted at north and south can are rewritten as 
\bel
&&\langle (\mathrm{Tr} \sigma )^M |_{\mathrm{N}}   \rangle_{\hbar}
=\sum_{ \{j \} } \oint_{x_a=-\lambda_{j_a}} \prod_{a=1}^n\frac{d x_a}{2\pi \sqrt{-1}}  
\tilde{Z}(x, \lambda,\lambda') Z ( z, x, \lambda, \lambda', \hbar) \non
&&~~~~~~~~~~~~~~~~ \times \left( \sum_{\mathbf{k} \in \mathbb{Z}^n_{ \ge 0}} ((-1)^{n-1} z)^{ \sum_{a=1}^{n} k_a} \prod_{a=1}^n (x_a -k_a \hbar)^M Z_{\Phi} (k, x,\lambda, -\hbar) \prod_{l=1}^s Z_{P_l} (k, x, \lambda', -\hbar) \right),
 \non
&&\langle (\mathrm{Tr} \sigma )^M |_{\mathrm{S}}   \rangle_{\hbar}
=\sum_{ \{j \} } \oint_{x_a=-\lambda_{j_a}} \prod_{a=1}^n\frac{d x_a}{2\pi \sqrt{-1}}  
\tilde{Z}(x, \lambda, \lambda') Z ( z, x, \lambda, \lambda', -\hbar) \non
&&~~~~~~~~~~~~~~~~ \times
\left( \sum_{\mathbf{k} \in \mathbb{Z}^n_{ \ge 0}} ((-1)^{n-1} z)^{ \sum_{a=1}^{n} k_a} \prod_{a=1}^n (x_a+k_a \hbar)^M Z_{\Phi} (k, x,\lambda , \hbar) \prod_{l=1}^s Z_{P_l} (k, x, \lambda', \hbar) \right),
 \non 
\ee
with
\bel
&&
\tilde{Z}(x,\lambda, \lambda')= \frac{\prod_{n \ge a >b \ge 1}(x_a-x_b)^2 \prod_{l=1}^s \prod_{\rho \in \Delta(R_l)}
 (\rho (x)+\lambda_l' ) }{  {\prod_{i=1}^m \prod_{a=1}^n} ( x_a +\lambda_i )},\\
&&Z (z, x, \lambda, \lambda', \hbar)
=\sum_{ \mathbf{k} \in \mathbb{Z}^n_{\ge 0} } ((-1)^{n-1} z)^{ \sum_{a=1}^{n} k_a} Z_{\Phi} ( k, x, \lambda, \hbar) 
\prod_{l=1}^s Z_{P_l} ( k,  x, \lambda', \hbar), 
\label{Zfunction}
\\ 
&& Z_{\Phi} (k, x, \lambda, \hbar) 
= \frac{{\prod_{n \ge a > b \ge 1}} 
\left(x_{a}-x_{b}+(k_a-k_b)\hbar \right)}
{ { \prod_{n \ge a > b \ge 1} }(x_{a}-x_{b}) 
{ \prod_{i=1}^{m} \prod_{a=1}^{n}\prod_{l=1}^{k_a}}
  (x_{a}+\lambda_i+l \hbar)  },  \\
&& Z_{P_{l}}  ( k, x, \lambda', \hbar)  = \prod_{\rho \in \Delta(R_l)} \prod_{j=1}^{-\rho (\mathbf{k})} (\rho (x)+\lambda_l'-j \hbar).
\ee
Here $\{ j\}=\{j_1, j_1, \cdots, j_n \} $ and $\sum_{\{ j \}}= \sum_{ 1 \le j_1 <j_2 < \cdots < j_n \le m }$.
Then the generating function of the correlation functions $\langle (\mathrm{Tr} \sigma )^M |_{\mathrm{N}} ( \mathrm{Tr} \sigma )^N |_{\mathrm{S}}  \rangle_{\hbar}$
can be written as 
\bel
\langle e^{\alpha \mathrm{Tr} \sigma } |_{\mathrm{N}} e^{\beta \mathrm{Tr} \sigma } |_{\mathrm{S}}  \rangle_{\hbar}
&&=\sum_{\{j \}} \oint_{x_a=-\lambda_{j_a}}  \prod_{a=1}^n\frac{d x_a}{2\pi \sqrt{-1}}  %Z_{1\mathchar `-\text{loop}}(\sig)
\tilde{Z}(x, \lambda, \lambda') \nonumber \\
&&\times e^{\alpha \sum_{a=1}^n  x_a } Z (e^{-\alpha \hbar} z, x, \lambda, \lambda', -\hbar)  
e^{\beta \sum_{a=1}^n  x_a } Z (e^{\beta \hbar } z, x, \lambda, \lambda', \hbar). 
\label{double1}
\ee
When the twisted masses are distinct, we can  explicitly perform the contour integrals in \eqref{double1} and  obtain the following 
vortex factorization form of the generating function.
\bel
\langle e^{\alpha \mathrm{Tr} \sigma } |_{\mathrm{N}} e^{\beta \mathrm{Tr} \sigma } |_{\mathrm{S}}  \rangle_{\hbar}
=  
\sum_{\{ j \}} \tilde{Z}_{\{ j \}}(\lambda, \lambda') 
e^{\alpha \sum_{a=1}^n  \lambda_{j_a} } Z_{\text{v}, \{ j \} } (e^{-\alpha \hbar} z,  \lambda, \lambda', -\hbar)  
e^{\beta \sum_{a=1}^n  \lambda_{j_a} } Z_{\text{v}, \{ j \} } (e^{\beta \hbar } z , \lambda, \lambda', \hbar). \non
\label{double2}
\ee
with
\bel
&&\tilde{Z}_{\{ j \}}(\lambda, \lambda')=\frac{{\prod_{n \ge a >b \ge 1}(\lambda_{j_a}-\lambda_{j_b})^2 \prod_{l=1}^s
 \prod_{\rho \in \Delta ( R_l)} (\rho (-\lambda_{\{j \} })+\lambda_l')}}{\prod_{a=1}^n \prod_{i=1, i \neq j_a}^m  (  \lambda_i -\lambda_{j_a} )}, \\
&& Z_{\text{v}, \{ j \} } ( z , \lambda, \lambda', \hbar)=Z (z, x, \lambda, \lambda', \hbar) \Big|_{x_a=-\lambda_{j_a}}.
\label{vortfac}
\ee
Here $\lambda_{\{ j \}}=\mathrm{diag} (\lambda_{j_1}, \lambda_{j_2}, \cdots, \lambda_{j_n})$.
From the view point of Higgs branch localization, $\tilde{Z}_{\{ j \}}(\lambda, \lambda')$ is interpreted as the 1-loop determinant and
the vortex partition function $Z_{\text{v}, \{ j \} } ( z , \lambda, \lambda', -\hbar) $ $(Z_{\text{v}, \{ j \} } ( z , \lambda, \lambda', \hbar)) $ 
is the point like vortex contribution on north (south) pole of $S^2$ at a root of Higgs branch specified by twisted masses $\lambda_{\{ j \}}$. 
 
Now we discuss the relation
between \eqref{omegaS2}, \eqref{double2} and Givental's work \cite{MR1354600,MR1408320}.
For clarity of exposition,
we restrict ourselves to the case
when the gauge group is $G=U(1)$,
the gauge charge of $\Phi^i$ is 1 for $i=1, \ldots, m$, and
the gauge charge of $P_l$ is $-q_l$ with $m=\sum_{l=1}^s q_l$.
The target space of the low-energy NLSM is
a CY complete intersection in $\mathbb{P}^{m-1}$.
Then \eqref{omegaS2} with $f(\sigma)=e^{\alpha \sigma}$ and $g(\sigma)=1$
is expressed as
\bel
 \langle e^{\alpha \sigma} |_{\mathrm{N}}   \rangle_{\hbar}
  = (-1)^s \sum_{k=0}^{\infty}  ( (-1)^m z  )^{ k}
 \oint  \frac{d x}{2\pi \sqrt{-1}}  e^{\alpha x} 
 \frac{\prod_{l=1}^s  \prod_{j=0}^{q_l k} (q_l x -\lambda_l' +j \hbar)}
 {\prod_{i=1}^m  \prod_{j=0}^{k}(x  +\lambda_i+j \hbar)},
\ee
which agrees with the function
\begin{align}
 \Phi^* =
  \frac{1}{2 \pi \sqrt{-1}}
  \oint e^{p(t-\tau)/\hbar}
   \sum_{d=0}^\infty e^{d \tau}
    \frac{\prod_{a=1}^r \prod_{m=0}^{l_a d} (l_a p - \lambda_a' - m \hbar)}
     {\prod_{i=0}^n \prod_{m=0}^d (p-\lambda_i-m \hbar)}
  d p
\end{align}
appearing in \cite[page~650]{MR1408320}
by setting 
$\alpha \hbar =-(\tau -t)$, and $\log ( (-1)^m z)=\tau$
up to change of signs of $\lambda_i$, $\hbar$ and overall sign.
The function $\Phi^*$ is
the generating function
of intersection numbers on the quasimap space,
and goes back to the generating function
\begin{align}
 \sum_{d=0}^\infty e^{dt} \int_{M_d} E_{d,l} e^{(t-\tau)(A+\omega/\hbar)}
\end{align}
in \cite[page~338]{MR1354600},
which is a regularized version of a `$\frac{\infty}{2}$-dimensional integral'
on the loop space.
The factorization
\begin{align}
 \Phi^* &= \sum_i \frac{\prod_a(l_a \lambda_i - \lambda_a')}
  {\prod_{j \ne i} (\lambda_i-\lambda_j)}
  e^{\lambda_i(t-\tau)/\hbar}
  Z_i^*(e^t,\hbar)
  Z_i^*(e^\tau,-\hbar),
    \\
 Z_i^* &= \sum_{d=0}^\infty q^d
  \frac{\prod_{a=1}^r \prod_{m=1}^{l_a d} (l_a \lambda_i - \lambda_a' + m \hbar)}
   {\prod_{\alpha=0}^n \prod_{m=1}^{d} (\lambda_i - \lambda_\alpha + m \hbar)},
\end{align}
which agrees with the Abelian case
\begin{align}
 \langle e^{\alpha \sigma } |_{\mathrm{N}} \rangle_{\hbar}
  &=
  \sum_{ j =1}^m \tilde{Z}_{ j }(\lambda, \lambda') 
   e^{\alpha   \lambda_{j} }
   Z_{\text{v},  j } (e^{-\alpha \hbar} z,  \lambda, \lambda' , -\hbar)  
   Z_{\text{v},  j  } ( z , \lambda, \lambda' , \hbar), \\
 \tilde{Z}_{ j }(\lambda, \lambda') &=
  \frac{{\prod_{l=1}^s ( q_l \lambda_j +\lambda_l')}}
  {\prod_{i=1, i \neq j}^m (\lambda_i - \lambda_j)}, \\
 Z_{\text{v},  j  } ( z , \lambda, \lambda', -\hbar) &=
  \sum_{k=0}^{\infty} ((-1)^m z)^k
   \frac{\prod_{l=1}^s \prod_{p=1}^{ q_l k} (-q_l \lambda_j -\lambda_l'- p \hbar)}
    {\prod_{i=1}^m \prod_{l=1}^k (\lambda_i - \lambda_j - l \hbar)}
\end{align}
of \eqref{double2},
also goes back to \cite[page~338]{MR1354600},
and is an important ingredient
in the proof of Givental's mirror theorem.
The factorization for toric complete intersections
is given in \cite[Proposition 6.2]{MR1653024}.
 
Givental's $I$-function $I(z, x,  \hbar)$
for the complete intersection in $\mathbb{P}^{m-1}$
is given by
\bel
I (z, x, \hbar)=z^{\frac{x}{\hbar}} Z ((-1)^m z, x,\lambda, \lambda', \hbar) \Big|_{\lambda_i=\lambda_l'=0}
=z^{\frac{x}{\hbar}}  \sum_{k=0}^{\infty }  z^k 
\frac{{ \prod_{l=1}^s \prod_{j=1}^{q_l k} } (q_l x+ l \hbar ) } 
{{ \prod_{l=1}^{k}}  (x+l \hbar)^{m} }. 
\ee
The period integrals $I_i (z)$ appear in the expansion of the $I$-function as 
\bel
 I (z, x, \hbar) = \sum_{i=0}^3 I_i(z) \left(\frac{ x }{\hbar} \right)^i
  \mod \left(\frac{x}{\hbar} \right)^4.  
\ee

When the gauge group is non-Abelian,
we expect that $Z (z, x, \lambda, \lambda', \hbar)$ is again related to Givental's $I$-function
$I( z, x , \hbar)$ of $X^{n,m}_{\oplus_{l=1}^s E^*_{l}}$ by
\bel
I ({z}, x, \hbar)=z^{\sum_{a=1}^n \frac{x_a} {\hbar}} Z ( (-1)^m z, x, \lambda, \lambda', \hbar)\Big|_{\lambda_i=\lambda_l'=0}. 
\label{Ifunction2}
\ee
Here the sign of $z$ is fixed by requiring the first instanton number to be positive. 
If the $I$-function for a Calabi--Yau 3-fold is expanded as 
\bel
 I (z, x, \hbar) = I_0 (z) + I_1 (z) \frac{\sum_{a=1}^n x_a}{\hbar}
  + O\left(\frac{1}{\hbar^2} \right),
\label{Ifunction}
\ee
then the mirror map is again given by the ratio of two periods as (\ref{mirrormap}).
If we define $Z_0(z)$ and $Z_1(z)$
as the first two coefficients of expansion
\bel
Z ((-1)^m z, x, \hbar)=Z ( (-1)^m z, x, \lambda, \lambda', \hbar)\Big|_{\lambda_i=\lambda_l'=0}=Z_0(z)+Z_1(z) \frac{\sum_{a=1}^n x_a}{\hbar}+O\left(\frac{1}{\hbar^2} \right). \non
\label{Zfunction0}
\ee
Then $I_0$ and $I_1$ are related to $Z_0$ and $Z_1$ by
\bel
 I_0(z)=Z_0(z), \quad {I_1(z)}=Z_0(z)\log(z)+{Z_1(z)}.
\ee

%%%%%%%%%%%%%%%%%%%%%%%%%%%%%%%%%%%%%%%%%
\section{Computation of  Gromov--Witten invariants}

In this section,
we compute the Yukawa couplings and
genus-zero Gromov--Witten invariants
of some examples of compact CY 3-folds in Grassmannians
which are obtained as complete intersections of equivariant vector bundles.

%%%%%%%%%%%%%%%%%%%%%%%%%%%%%%%%%%%%%%%%%%
\begin{table}[thb]
\begin{center}
  \begin{tabular}{|c|c  c c| }
  \hline            &$\phi_{i}, (i=1,\cdots, 7)$    &       $p_1$ & $p_l, (l=2,\cdots, 5)$   \\
\hline $U(2)_G$    & $\mathbf{2}$             & $\mathrm{Sym}^2 {\mathbf{2}}^*$  & $\mathrm{det}^{-1}$                \\  
         $U(1)_R$     & $0$     &  $2$   & $2$      \\  
  \hline
  \end{tabular}
\end{center}
\caption{The charge assignment for lowest component scalars
in the chiral multiplets of equivariant A-twisted GLSM
for $X^{2,7}_{\mathrm{Sym}^2 \mS^* \oplus \cO(1)^{\oplus 4}}$}
\label{tab1}
\end{table}

Our first example is
$X^{2,7}_{\mathrm{Sym}^2 \mathcal{\mS}^* \oplus \cO(1)^{\oplus 4}}$,
which is a complete intersection CY 3-fold of
$\mathrm{Sym}^2 \mathcal{\mS}^*$ and
four copies of $\cO(1)$
in $\Gr (2,7)$.
The gauge group is $U(2)$, and there are
seven fundamental chiral multiplets $\Phi_i, (i=1,\cdots,7)$,
one chiral multiplet $P_1$
in the gauge representation $\mathrm{Sym}^2 \mathbf{2}^* $, and 
four chiral multiplet $P_i (i=2,\cdots, 5)$
in the gauge representation $\mathrm{det}^{-1}$.
The charge assignment for the lowest component scalars
in the chiral multiplets is listed in Table \ref{tab1}. 
The superpotential is given by
\bel
 W
  = \sum_{l=1}^5 P_l G_{l}(\Phi)
  = P^{(ab)}_1 S_{(i j)} \Phi^{(i}_{(a} \Phi^{j)}_{b)}
   +\sum_{l=2}^5   P_l A_{[ij]}^l \ep^{ab} \Phi^{[i}_{a} \Phi^{j]}_{b}.
\ee
Here $a,b$ denote color indices, and the notations $(ij)$ and $[i,j]$ show
that the indices $i$ and $j$ are symmetric and anti-symmetric
under the permutation respectively.
The F-term equation for $P_l$ gives
\bel
&& A_{[ij]}^l  \ep^{ab} \phi^{[i}_{a} \phi^{j]}_{b}=0,
 \quad S_{(i j)}\phi^{(i}_{(a} \phi^{j)}_{b)}=0.
\ee

Let us compute the mirror Yukawa coupling and instanton numbers.
The set of weights of the chiral multiplet $P_l$
evaluated at $\mathrm{diag}(\sigma_1, \sigma_2)$ 
is given by
\bel
 \begin{matrix} 
 &\{-2\sigma_1, -\sigma_1-\sigma_2, -2\sigma_2 \} & \quad  \text{ for } P_1, \\
 &\{-\sigma_1-\sigma_2 \} & \quad   \text{ for } ~ P_2, \ldots, P_5.
\end{matrix}
\ee
Hence the contributions from the one-loop determinants
of the chiral multiplets are
\bel
Z^{\mathrm{chiral}}_{\mathbf{k}}(\sig,\hbar=0) =
\begin{cases}
\prod_{a=1}^2 \sigma^{-k_a-1}_a, & \text{for } \Phi_i, \\
(-\sig_1-\sig_2)^{k_1+k_2+1} \prod_{a=1}^2(-2\sig_a)^{2k_a+1}
 & \text{for } P_1 \\
(-\sig_1-\sig_2)^{k_1+k_2+1}  & \text{for } P_2, \ldots, P_5.
\end{cases}
\ee
From \eqref{omegazero},
the Yukawa coupling of this model
with the sign change $z \to -z$
is given by
\bel
\langle (\mathrm{Tr} \sigma )^3 \rangle_{\hbar=0} &&=\frac{1}{2} \sum_{\mathbf{k} \in \mathbb{Z}^2_{\ge 0}}
\oint_{\sigma=0} \prod_{a=1}^2 \frac{d \sig_a}{2\pi \sqrt{-1}} z^{k_1+k_2} (\sigma_1+\sigma_2)^3 (\sigma_1- \sig_2)^2 \non
&& ~~~~ \times \frac{(-\sig_1-\sig_2)^{5k_1+5k_2+5} (-2\sig_1)^{2k_1+1} (-2\sig_2)^{2k_2+1}}{\prod_{a=1}^2 \sigma^{7k_a+7}_a} \non
&&=\sum_{k=0}^{\infty} \sum_{m=0}^k (-1)^{k+1} 2^{2k+2} z^k  \left( {{5k+6} \choose{5m+1}}
-{{5k+6} \choose {5m+3}} \right)
 \non 
&&=\frac{8(7+6z)}{(1+4z)(1-44z-16z^2)}.
\ee
%%%%%%%%%%%%%%%%

Next we compute the Yukawa coupling
in the flat coordinate.
\eqref{Zfunction0} is written in this model as
\bel
&&Z(-z, x, \hbar)=\sum_{\mathbf{k} \in \mathbb{Z}^2_{\ge 0}}  
(-z)^{ k_1+k_2} \frac{  
\left(x_{2}-x_{1}+(k_2-k_1)\hbar \right)  
\displaystyle{\prod_{l=1}^{k_1+k_2} (x_1+x_2+l\hbar)^5 \prod_{a=1}^2 \prod_{l=1}^{2 k_a} 
(2x_a+l\hbar) }
 }
{ (x_{2}-x_{1}) 
\displaystyle{ \prod_{a=1}^{2}\prod_{l=1}^{k_a}}
  (x_{a}+l \hbar)^{7}  }. \non 
\ee
From this equation, the series $Z_0(z)$ and $Z_1(z)$ can be read off as  
\bel
&&Z_0 (z)=1 + 4 z + 64 z^2 + 1408 z^3 + 37216 z^4 + 1093504 z^5+\cdots \\
&&Z_1(z)= 10 z + 189 z^2 + \frac{13528}{3} z^3+ \frac{744743}{6} z^4 + \frac{11218906}{3} z^5+\cdots
\ee
 (\ref{mirrormap}) is solved  recursively as 
\bel
z(q)=q - 10 q^2 + q^3 + 20 q^4 - 2412 q^5+\cdots.
\ee
Then, we obtain the  Yukawa coupling in the flat coordinate: 
\bel
K_{ttt} =n_0+\sum_{d=1}^{\infty} n_d d^3 \frac{q^d}{1-q^d}, 
\ee
with
\bel
n_0=56,~n_1=160,~n_2=758,~n_3=5824,~n_4=65540,~n_5=884064,\cdots.
\label{GW1}
\ee
(\ref{GW1}) reproduce the correct triple intersection number
and the genus-zero instanton numbers
for No.212 in the Calabi--Yau date base \cite{Straten}.
Genus-zero Gromov--Witten invariants
for other $U(2)$ GLSMs are listed in the table \ref{GWtable1}.
The Calabi--Yau 3-fold
$X^{2,6}_{\mS^* \otimes \cO(1)\oplus \cO(1)^{\oplus 3}}$
is known by \cite{Inoue1}
to be deformation-equivalent
to a complete intersection Calabi--Yau 3-fold
in a minuscule Schubert variety
introduced by Miura \cite{Miura1}.
Its genus-zero Gromov--Witten invariants are also
computed in \cite{Miura1}.
Gromov--Witten invariants for Miura's Calabi--Yau 3-fold was also computed in \cite{Gerhardus:2015sla} by  physical $S^2$ partition function method \cite{Jockers:2012dk}.

\begin{table}
\begin{center}
\begin{tabular}{|c|r|r|r|r|}
\hline
& $X^{2,5}_{\mS^{*} \otimes \cO(1) \oplus \cO(2)}$ 
& $X^{2,6}_{\mS^{*} \otimes \cO(1)\oplus \cO(1)^{\oplus 3}}$ 
& $X^{2,6}_{\mathrm{Sym}^2 \mS^{*} \otimes \cO(1) \oplus \cO(2)}$
& $X^{2,6}_{\mathrm{Sym}^2 \mS^{*} \oplus \cS^* \otimes \cO(1)}$  \\
   \hline
$n_0$  & 24           &  33   & 40 & 48 \\
$n_1 $ & 336         & 252 & 160 &  112\\
$n_2 $ &3636         & 1854 & 1560 &  1102 \\
$n_3 $ &83392       & 27156 &  14560  & 7104 \\
$n_4 $ &2727936   & 567063 & 272000 & 98892 \\
$n_5 $ &109897632& 14514039 &5299328  & 1389664\\
\hline
\end{tabular}
\caption{The triple intersection number and genus zero instanton numbers for CY 3-folds in $\Gr(2,m)$.}
\label{GWtable1}
\end{center}
\end{table}

%%%%%%%%%%%%%%%%%%%%%%%%%%%%%%%%%%%%
\begin{table}[htb]
\begin{center}
  \begin{tabular}{|c|  c c| }
  \hline            &$\phi_{i}, (i=1,\cdots, 5)$    &        $p$   \\
\hline $U(3)_G$    & $\mathbf{3} $             & $(\Lambda^2 {\mathbf{3}}^*) \otimes \mathrm{det}^{-1} $                 \\  
         $U(1)_R$     & $0$      & $2$      \\  
  \hline
  \end{tabular}
\end{center}
\caption{The matter contents of equivariant A-twisted GLSM for $ X^{3,5}_{(\Lambda^2\mS^*) \otimes \cO(1)}$. }
\label{tab3}
\end{table}
Our second example is
$ X^{3,5}_{(\Lambda^2\mS^*) \otimes \cO(1)}$,
which is known by \cite{Inoue1} to be deformation-equivalent
to the complete intersection of two copies of $\Gr(2,5)$
in $\mathbb{P}^9$.
This is a $U(3)$ GLSM with five fundamental chiral multiplets and
one chiral multiplet $P$ in the gauge representation
$(\Lambda^2 {\mathbf{3}}^*) \otimes \mathrm{det}^{-1} $
as shown in Table~\ref{tab3}.
The set of weights $\rho (\sigma)$ for the chiral multiplet $P$ is given by
\bel
 \{-(\sig_1+2\sig_2+2\sig_3),~ -(2\sig_1+\sig_2+2\sig_3),~ -(2\sig_1+2\sig_2+\sig_3)  \}.
\ee
The superpotential is 
\bel
W=A_{[i_1  i_2, i_3] [  i_4 i_5]}   \epsilon^{a_1 a_2 a_3}  P^{ [b c]  }
 \Phi_{a_1}^{[i_1} \Phi_{a_2}^{i_2} \Phi_{a_3}^{i_3]} \Phi_{[b}^{[i_4} \Phi_{c]}^{i_5]} .
\ee
From (\ref{omegazero}),
the Yukawa coupling of this model with the sign change $z \to -z$
is given by
\bel
\langle (\mathrm{Tr} \sigma )^3\rangle_{\hbar=0}&&=
\frac{1}{3!} \sum_{\mathbf{k} \in \mathbb{Z}^3_{\ge 0}} (-z)^{\sum_{a=1}^3 k_a} \oint_{\sigma=0} \prod_{a=1}^{3} \frac{d\sigma_a}{2\pi \sqrt{-1}} 
\left(\sum_{a=1}^3 \sigma_a \right)^3 \non
&& ~~~~~~~ \times  \prod_{a < b} (\sigma_a -\sigma_b)^2 
\prod_{a=1}^3 \frac{ (\sig_a-2\sum_{b=1}^3 \sig_b  )^{2\sum_{b=1}^3 k_b -k_a  +1 }}{\sig^{5k_a+5}_a}  \non
&&=25( 1+ 121 z +14884 z^2+1830609 z^3 +225150025 z^4 +27691622464 z^5) +\cdots.
\non 
\ee
The function (\ref{Zfunction0})  for this model is given by
\bel
&&Z(-z,x,\hbar)=\sum_{\mathbf{k} \in \mathbb{Z}^3_{\ge 0}} 
z^{ \sum_{a=1}^3 k_a} \frac{ 
\displaystyle{\prod_{a >b}} \left(x_{a}-x_{b}+(k_a-k_b)\hbar \right)
  \displaystyle{\prod_{a=1}^3  \prod_{j=1}^{2\sum_{b=1}^3 k_b - k_a} }  (2\sum_{b=1}^3 x_b -x_a + j \hbar )  }
{ \displaystyle{\prod_{a >b}} \left(x_{a}-x_{b}\right)
\displaystyle{ \prod_{a=1}^{3}\prod_{j=1}^{k_a}}
  (x_{a}+j \hbar)^{5}  }. \non
\ee
From this equation, we obtain the first two coefficients  as
\bel
&&Z_0 (z)=1 + 9 z + 361 z^2 + 21609 z^3 + 1565001 z^4+126630009 z^5+\cdots, \\
&&Z_1(z)=30 z + 1425 z^2 + 90895 z^3 + \frac{13604625}{2} z^4+\frac{1123000637}{2} z^5 +\cdots.
\ee
The complex structure moduli is expressed as function of $q$ as
\bel
z(q)=q - 30 q^2 + 195 q^3 - 3070 q^4 - 99495 q^5+\cdots.
\ee
Then the expected genus zero instanton numbers are 
\bel
n_0=25,~n_1=325,~n_2=3200,~n_3=66250,~n_4=1985000,~n_5=73034875, \cdots. \nonumber \\
\label{GW}
\ee
These values reproduce the Gromov--Witten invariants
of the complete intersection of two copies of $\Gr(2,5)$
in $\mathbb{P}^9$
calculated in \cite{Miura}.
Gromov--Witten invariants for some other CY 3-folds
in Grassmannian $\Gr(n,m)$
is listed in Tables \ref{GWtable2}, \ref{GWtable3} and \ref{GWtable4}.
The manifold
$X^{3,8}_{(\mathrm{Sym}^2 \mS^{*})^{ \oplus 2}}$
in Table \ref{GWtable2}
is an Abelian 3-fold,
so that its Gromov--Witten invariants are zero
except at degree 0.

\begin{table}[phtb]
\begin{center}
\begin{tabular}{|c|r|r|r|}
\hline
  &  
 $X^{3,6}_{\Lambda^2 \mS^{*}  \oplus \cO(1)^{\oplus 2} \oplus  \cO(2)}$ & $X^{3,6}_{\mS^{*}\otimes \cO(1) \oplus \Lambda^2 \mS^{*}}$ &   $X^{3,7}_{\mathrm{Sym}^2 \mS^{*} \oplus \cO(1)^{\oplus 3}}$ \\ \hline
$n_0$         & 32 & 42 &128 \\
$n_1 $    &  256 &  210 & 0\\
$n_2 $    & 2016 & 1176 &  4096 \\
$n_3 $     &  32000&13104 & 0\\
$n_4 $     &  709904&  201936&  9280 \\
$n_5 $     &  19397376&  3824016 &  0  \\
\hline
  & $X^{3,7}_{(\Lambda^2 \mS^{*})^{ \oplus 2}\oplus \cO(1)^{\oplus 3}}$ &
$X^{3,8}_{(\Lambda^2 \mS^{*})^{ \oplus 4}}$ & $X^{3,8}_{(\mathrm{Sym}^2 \mS^{*})^{ \oplus 2}}$\\ \hline
$n_0$           & 61 & 92 & 384 \\
$n_1 $     & 163 & 140 & 0\\
$n_2 $   &  630&  328 & 0\\
$n_3 $     & 4795&1872 & 0 \\
$n_4 $             & 48422 & 12280 &0\\
$n_5 $         &   599809   &  100728 & 0  \\
\hline
\end{tabular}
\caption{The triple intersection number and instanton numbers for CY 3-folds in $\Gr(3,m)$. }
\label{GWtable2}
\end{center}
\end{table}
\begin{table}[phtb]
\begin{center}
\begin{tabular}{|c|r|r|r|}
\hline
  & $X^{4,6}_{ \mS^{*} \otimes \cO(1)  \oplus \cO(1)}$  
  & $X^{4,7}_{\Lambda^2 \mS^{*} \oplus \cO(1)^{\oplus 2} \oplus \cO(2) }$
  & $X^{4,7}_{(\Lambda^3 \mS^{*})^{\oplus 2} \oplus \cO(1) }$ \\ \hline
$n_0$   & 42  &    32  & 72 \\
$n_1 $   & 196 & 256   & 136 \\
$n_2 $  &  1225&  2016 & 508 \\
$n_3 $   &  12740 & 32000 & 3088 \\
$n_4 $  & 198058  & 709904 &   25342  \\
\hline
\end{tabular}
\caption{The triple intersection number and instanton numbers for CY 3-folds in $\Gr(4,m)$.}
\label{GWtable3}
\end{center}
\end{table}
\begin{table}[phtb]
\begin{center}
\begin{tabular}{|c|r|r|}
\hline
  & $X^{5,7}_{ \Lambda^4 \mS^{*}  \oplus \cO(1) \oplus \cO(2)}$ & $X^{6,8}_{ \Lambda^5 \mS^{*}   \oplus \cO(1)^{\oplus 3}}$ 
    \\ \hline
$n_0$    &36  & 57  \\
$n_1 $  & 216  & 147 \\
$n_2 $   &1674 & 756 \\
$n_3 $   & 21888 & 5283 \\
\hline
\end{tabular}
\caption{The triple intersection number and instanton numbers for CY 3-folds in $\Gr(5,7)$ and $\Gr(6,8)$.}
\label{GWtable4}
\end{center}
\end{table}

%\newpage
\section{Seiberg like description of the mirror Yukawa coupling} 

In this section,
we study dual $U(m-n)$ GLSM description of Yukawa couplings
of $U(n)$ GLSM with $m$ fundamental chiral
and chiral multiplet $P_l, (l=1,\cdots, s)$
in several examples.
We start with the case where all $P_l$ belong
to the gauge representation $\mathrm{det}^{-q_l}$
and the target space of low energy NLSM is
$X^{n, m}_{\oplus_{l=1}^s \cO(q_l) }$
in $\Gr(n,m)$.
The Seiberg-like duality of this model is studied in \cite{Hori:2006dk}.
The CY 3-fold $X^{n, m}_{\oplus_{l=1}^s \cO(q_l) }$ 
is isomorphic to a CY 3-fold
$X^{m-n,m}_{\oplus_{l=1}^s \cO(q_l)}$
in $\Gr(m-n,m)$ .
Then, in the dual side,
the GLSM is $U(m-n)$ gauge group
with $m$ fundamental chiral multiplets
and chiral multiplet $P_l$ in the $\mathrm{det}^{-q_l}$ representation
for $l=1,\cdots, s$.
For example, one has
$
 X^{2, 6}_{ \cO(1)^{\oplus 4} \oplus \cO(2)}
  \simeq X^{4,6}_{ \cO(1)^{\oplus 4} \oplus \cO(2)}.
$
In $U(2)$ GLSM description,
the Yukawa coupling of
$X^{2, 6}_{ \cO(1)^{\oplus 4} \oplus \cO(2)}$
is given by
\bel
\langle (\mathrm{Tr} \sigma)^3 \rangle_{\hbar=0}
&&=\frac{1}{2} \sum_{\mathbf{k} \in \mathbb{Z}^2_{\ge 0}}  (-z)^{\sum_{a=1}^2 k_a}
\oint_{\sigma=0} \prod_{a=1}^2 \frac{d \sig_a}{2\pi \sqrt{-1}}    \left(\sum_{a=1}^2 \sigma_a \right)^{3}
\non &&~~~~~~~~ \times
 \left(\sigma_1-\sigma_2 \right)^2  
\frac{(-\sig_1 -\sigma_2)^{4 (k_1+k_2)+4} (-2\sig_1 -2\sigma_2)^{2(k_1+k_2)+1} }
{\prod_{a=1}^2 \sigma_a^{6(k_a+1)}} \non
&&= 28 (1 + 104 z + 11248 z^2 + 1214720 z^3)+ \cdots.
\label{Yukawa46}
\ee
In the dual U(4) GLSM description,
the Yukawa coupling of
$X^{4,6}_{ \cO(1)^{\oplus 4} \oplus \cO(2)}$
is given by
\bel
\langle (\mathrm{Tr} \sigma)^3 \rangle_{\hbar=0}
&&=-\frac{1}{24} \sum_{\mathbf{k} \in \mathbb{Z}^4_{\ge 0}}  (-z)^{\sum_{a=1}^4 k_a}
\oint_{\sigma=0} \prod_{a=1}^4 \frac{d \sig_a}{2\pi \sqrt{-1}}    \left(\sum_{a=1}^4 \sigma_a \right)^{3}
\non &&~~~~~~~~~ \times
\prod_{1 \le a < b \le 4} \left(\sigma_a-\sigma_b \right)^2  
\frac{(-\sum_{a=1}^4 \sig_a )^{4(\sum_{a=1}^4 k_a+1)} (-2\sum_{a=1}^4 \sig_a )^{(\sum_{a=1}^4 2 k_a+1)} }
{\prod_{a=1}^4 \sigma_a^{6(k_a+1)}} \non
&&=  28 (1 + 104 z + 11248 z^2 + 1214720 z^3)+\cdots.
\label{Yukawadual46}
\ee
From (\ref{Yukawa46}) and (\ref{Yukawadual46}),
we find that two A-twisted GLSMs give the same Yukawa coupling.

Next, we study the dual $U(m-n)$ GLSM description
of the Yukawa coupling
for a $U(n)$ GLSM
with $m$ fundamental chiral multiplets,
one chiral multiplet $P_1$
in the gauge representation ${\mathbf{n}}^* \otimes \mathrm{det}^{-1}$ 
and $P_2, \ldots, P_s$
in the gauge representation $\mathrm{det}^{-q_l}$,
which flow to NLSM with target space
$X^{n,m}_{\cS^* \otimes \cO(1)  \oplus_{l=2}^s \cO(q_l)}$.
This target space is a complete intersection of
$\cS^* \otimes \cO(1) $ and
$\cO(q_l)$ for $l=2,\cdots, s$ in $\Gr(n,m)$. 

To find the dual description,
let us first consider the dual $U(m-n)$ gauge theory description
of $U(n)$ with $m$ fundamental chiral multiplet $\Phi_i$
and an anti-fundamental chiral multiplet $P_1$ \cite{Jia:2014ffa}. 
In the $U(n)$ GLSM, an anti-fundamental chiral multiplet define the fiber
of the universal subbundle $\cS$ on $\Gr(n,m)$.
The universal subbundle $\cS  \to \Gr(n,m)$ is mapped
to the dual of the universal quotient bundle $\cQ^{*} \to \Gr(m-n,m)$. Then $\cQ^{*}  \to \Gr(m-n,m)$ can be realized by a $U(m-n)$ gauge theory
with $m$ fundamental chiral multiplets,
$m$ mesonic chiral fields $M_1, \ldots, M_m$
and a chiral multiplet $\tilde{\Phi}$ in the  representation $ (\mathbf{m-n})^*$
with the superpotential $W=\sum_{i=1}^m M_i \tilde{\Phi} \Phi_i$.
The F-term equation gives $\cQ^*$ on $\Gr(m-n,m)$. 

In our case,
the anti-fundamental chiral multiplet $P_1$ is modified
to chiral multiplet in the gauge representation $\mathbf{n}^* \otimes \mathrm{det}^{-1}$.
With the D-term equation,
an anti-fundamental chiral multiplet defines the fiber
of the vector bundle $\cS \otimes \cO(-1)$.
The bundle $\cS \otimes \cO(-1) \to \Gr(n,m)$ is mapped
to the tensor product $\cQ^{*} \otimes \cO(-1)$
of the dual of the universal quotient bundle and
the inverse of the determinant line bundle
on $\Gr(m-n,m)$.
Similarly, the bundle $\cQ^{*} \otimes \cO(-1)$ on $\Gr(m-n,m)$
can be realized by a $U(m-n)$ gauge theory
with $m$ fundamental chiral multiplet,
$m$ meson like chiral fields $M_1, \ldots, M_m$
in the representation $\mathrm{det}^{-1} $ of $U(m-n)$
and a chiral multiplet $\tilde{\Phi}$
in the $ (\mathbf{m-n})^* \otimes \mathrm{det}$ representation
with the superpotential $W=\sum_{i=1}^m M_i \tilde{\Phi} \Phi_i$.
The F-term equation gives  $\cQ^* \otimes \cO(-1)$ on $\Gr(m-n,m)$.
We expect that the matter context of $U(m-n)$ GLSM is
$m$ fundamental chiral multiplets $\Phi$,
$m$ chiral multiplets $M_i$ in the  representation $\mathrm{det}^{-1}$,
a chiral multiplet $\tilde{\Phi}$ in the representation $(\mathbf{m-n})^* \otimes \mathrm{det}$
and chiral multiplets $P_2, \ldots, P_s$
in the  representation $\mathrm{det}^{-q_l}$
with the superpotential
$W=\sum_{i=1}^mM_i \tilde{\Phi} \Phi_i+\sum_{l=2}^s P_{l} G_l (\Phi)$.  

\begin{table}[htb]
\begin{center}
  \begin{tabular}{|c|c  c c| }
  \hline            &$\phi_{i}, (i=1,\cdots, 6)$    &       $p_1$ & $p_2$   \\
\hline $U(4)_G$    & $\mathbf{4}$             & $ {\mathbf{4}}^* \otimes \mathrm{det}^{-1}$  & $\mathrm{det}^{-1}$                \\  
         $U(1)_R$     & $0$     &  $2$   & $2$      \\  
  \hline
  \end{tabular}
\end{center}
\caption{Field contents of $U(4)$ GLSM for $X^{4,6}_{ \mS^* \otimes \cO(1) \oplus \cO(1)}$}
\label{origial}
\end{table} 

We compute the Yukawa coupling in both sides and see the agreement.
We first consider the $U(4)$ GLSM description of
$X^{4,6}_{\cS^* \otimes \cO(1) \oplus \cO(1)}$.
The field content is listed in Table~\ref{origial}.
The Yukawa coupling is given by
\bel
\langle (\mathrm{Tr} \sigma)^3 \rangle_{\hbar=0}
&&=-\frac{1}{24} \sum_{\mathbf{k} \in \mathbb{Z}^4_{\ge 0}}  (-z)^{\sum_{a=1}^4 k_a}
\oint_{\sigma=0} \prod_{a=1}^4 \frac{d \sig_a}{2\pi \sqrt{-1}}    \left(\sum_{a=1}^4 \sigma_a \right)^{3}
\non && \times
\prod_{1 \le a < b \le 4} \left(\sigma_a-\sigma_b \right)^2  
 \frac{\prod_{a=1}^4(-\sigma_a -\sum_{b=1}^4 \sig_b )^{k_a+\sum_{b=1}^4 k_a+1} 
( -\sum_{b=1}^4 \sig_b )^{\sum_{b=1}^4 k_a+1} }{
\prod_{a=1}^4 \sigma_a^{6(k_a+1)}} \non
&&= 14 (3 + 170 z + 10557 z^2 + 650876 z^3 + 40150735 z^4)
+\cdots.
\label{Yukawas46}
\ee

\begin{table}[htb]
\begin{center}
  \begin{tabular}{|c|  c c c c| }
  \hline            &$\phi_{i}$    &        ${M}_i, (i=1,\cdots, 5)$  & $p_1$ & $\tilde{\phi}$ \\
\hline $U(2)_G$    & $\mathbf{2}$             & $\mathrm{det}^{-1}$ &   $\mathrm{det}^{-1}$ & $ {\mathbf{2}}^* \otimes \mathrm{det}  $     \\ 
         $U(1)_R$     & $0$      & $2$   & $2$ & $0$ \\  
  \hline
  \end{tabular}
\end{center}
\caption{Field contents of dual $U(2)$ GLSM for $X^{4,6}_{\cS^* \otimes \cO(1) \oplus \cO(1)}$}
\label{tab5}
\end{table}

The matter content of the dual $U(2)$ description of
$X^{4,6}_{\cS^* \otimes \cO(1) \oplus \cO(1)}$
is given in Table~\ref{tab5}.
The Yukawa coupling is given by
\bel
\langle (\mathrm{Tr} \sigma)^3 \rangle_{\hbar=0}
&&=\frac{1}{2} \sum_{\mathbf{k} \in \mathbb{Z}^2_{\ge 0}}  (-z)^{\sum_{a=1}^2 k_a}
\oint_{\sigma=0} \prod_{a=1}^2 \frac{d \sig_a}{2\pi \sqrt{-1}}    \left(\sum_{a=1}^2 \sigma_a \right)^{3}
\non &&~~~~~~~~~~~~~~~~ \times
 \left(\sigma_1-\sigma_2 \right)^2  
\frac{(-\sig_1 -\sigma_2)^{7(k_1+k_2)+7} }
{\prod_{a=1}^2 \sigma_a^{7(k_a+1)}} 
\non
\label{dualYukawa27}
&&= 14 (3 - 170 z + 10557 z^2 - 650876 z^3 + 40150735 z^4
)+ \cdots.
\label{dualYukawa27s}
\ee
We find (\ref{dualYukawa27s}) agrees with (\ref{Yukawas46})
up to the change $z \to -z$ of signs.
Note that (\ref{dualYukawa27}) has the same Yukawa coupling
as $X^{2,7}_{\cO(1)^{\oplus 7}}$,
which is known by \cite{Inoue1}
to be deformation-equivalent to
$X^{4,6}_{\cS^* \otimes \cO(1) \oplus \cO(1)}$ 

\begin{table}[htb]
\begin{center}
  \begin{tabular}{|c|c  c c| }
  \hline            &$\phi_{i}, (i=1,\cdots, 5)$    &       $p_1$ & $p_2$   \\
\hline $U(2)_G$    & $\mathbf{2}$             & $ {\mathbf{2}}^* \otimes \mathrm{det}^{-1}$  & $\mathrm{det}^{-2}$                \\  
         $U(1)_R$     & $0$     &  $2$   & $2$      \\  
  \hline
  \end{tabular}
\end{center}
\caption{Field content of GLSM for $X^{2,5}_{ \mS^* \otimes \cO(1) \oplus \cO(2)}$}
\label{tab25}
\end{table}

Next we study the dual $U(3)$ description of
$X^{2,5}_{\cS^* \otimes \cO(1) \oplus \cO(2)}$.
The field content of $U(2)$ GLSM is given in Table~\ref{tab25}.
In the original $U(2)$ GLSM,
the Yukawa coupling of
$X^{2,5}_{\cS^* \otimes \cO(1) \oplus \cO(2)}$
is given by
\bel
\langle (\mathrm{Tr} \sigma)^3 \rangle_{\hbar=0}
&&=\frac{1}{2} \sum_{\mathbf{k} \in \mathbb{Z}^2_{\ge 0}}  (-z)^{\sum_{a=1}^2 k_a}
\oint_{\sigma=0} \prod_{a=1}^2 \frac{d \sig_a}{2\pi \sqrt{-1}}    \left(\sum_{a=1}^2 \sigma_a \right)^{3}  \left(\sigma_1-\sigma_2 \right)^2  
\non &&~~~~~~~~~~~~~~~~ \times
\frac{\prod_{a=1}^2(-\sig_a-\sig_1 -\sigma_2)^{k_a+k_1+k_2+1} (-2\sig_1 -2\sigma_2)^{2(k_1+k_2)+1} }
{\prod_{a=1}^2 \sigma_a^{5(k_a+1)}} 
\non
&&=24 (1 - 136 z + 18480 z^2 - 2511104 z^3 + 341214464 z^4)+\cdots
\label{oriYukawa25}
\ee 

From our observation,
we expect that the matter content of $U(3)$ GLSM is given
by Table~\ref{tab6}
with the superpotential
$W=\sum_{i=1}^5 {M}_i \tilde{\Phi} \Phi_{i} + P_1 G(\Phi)$.
Here $G(\Phi)$ is a homogeneous polynomial of degree 4.

\begin{table}[htb]
\begin{center}
  \begin{tabular}{|c|  c c c c| }
  \hline            &$\phi_{i}$    &        ${M}_i, (i=1,\cdots, 5)$  & $p_1$ & $\tilde{\phi}$ \\
\hline $U(3)_G$    & $\mathbf{3}$             & $\mathrm{det}^{-1}$ &   $\mathrm{det}^{-2}$ & $  {\mathbf{3}}^* \otimes \mathrm{det}$     \\ 
         $U(1)_R$     & $0$      & $2$   & $2$ & $0$ \\  
  \hline
  \end{tabular}
\end{center}
\caption{Field contents of dual $U(3)$ GLSM for $X^{2,5}_{\cS^* \otimes \cO(1) \oplus \cO(2)}$}
\label{tab6}
\end{table}

The Yukawa coupling on the dual side is 
\bel
\langle (\mathrm{Tr} \sigma)^3 \rangle_{\hbar=0}&&=\frac{1}{6} \sum_{\mathbf{k} \in \mathbb{Z}^n} z^{\sum_{a=1}^n k_a} 
\sum_{\sigma_*} 
{{\rm JK \mathchar `-Res}(\mathbf{Q} (\sigma_*), \eta)} \non
&&\times \left( \sum_{a=1}^3 \sig_a \right)^3 \prod_{a < b} (\sigma_a - \sigma_b )^2 
 \frac{(-\sum_{a=1}^3 \sigma_a)^{5\sum_{a=1}^3 k_a +5}  (-2 \sum_{a=1}^3 \sigma_a)^{2\sum_{a=1}^3 k_a +1}}{\prod_{a <b} (\sig_a+\sig_b)^{k_a +k_b+1} 
\prod_{a=1}^3 \sig^{5k_a+5}_a} d^3 \sigma \non
\label{dual25} 
&&=: \sum_{\sigma_*}
{{\rm JK \mathchar `-Res}(\mathbf{Q} (\sigma_*), \eta)}  \omega.
\ee
(\ref{dual25}) is a degenerate case and
we use a constructive definition of Jeffrey-Kirwan residue operation
\cite{Benini:2013xpa}.   
The singular hyperplanes
$H_{a b}=\{  \sigma_a +\sigma_b=0 \}$ and $H_a=\{ \sigma_a=0 \}$
meet at the origin $\sigma_*=0$. 
The Jeffrey-Kirwan operation is not defined
in the physical choice of vector $\eta=(1,1,1)$,
and we slightly shift $\eta$ inside the geometric phase.
For example,
we can take $\eta=(1,1+\varepsilon,1-\varepsilon), \varepsilon < 1$. 
Then eight flags will contribute to the residue operation.
But the iterated residue for six of them gives zero
in an order-by-order computation.
The following iterated residues give non-zero contributions;
\bel
{\underset{\sigma_3=0}{\mathrm{Res}}} ~ {\underset{\sigma_1=0}{\mathrm{Res}}} ~ {\underset{\sigma_2=0}{\mathrm{Res}}} ~ \omega
=\frac{8}{3} (9 - 1216 z + 171232 z^2 - 19353984 z^3+5393384448 z^4)+\cdots, \non
\ee
and
\bel
{\underset{\tilde{\sigma}_3=0}{\mathrm{Res}}} ~ {\underset{\tilde{\sigma}_2=0}{\mathrm{Res}}} ~ {\underset{\tilde{\sigma}_1=0}{\mathrm{Res}}} ~ \omega
=-\frac{64}{3} (z + 614 z^2 + 405744 z^3+290306784 z^4)+\cdots, 
\ee
with $\tilde{\sigma}_1:=\sigma_1+\sigma_2, \tilde{\sigma}_2:=\sigma_2+\sigma_3, \tilde{\sigma}_3:=\sigma_1+\sigma_3$.
Then the Yukawa coupling is
\bel
\langle (\mathrm{Tr} \sigma)^3 \rangle_{\hbar=0}&&={\underset{\sigma_3=0}{\mathrm{Res}}} ~ {\underset{\sigma_1=0}{\mathrm{Res}}} ~ {\underset{\sigma_2=0}{\mathrm{Res}}} ~ \omega+
{\underset{\tilde{\sigma}_3=0}{\mathrm{Res}}} ~ {\underset{\tilde{\sigma}_2=0}{\mathrm{Res}}} ~ {\underset{\tilde{\sigma}_1=0}{\mathrm{Res}}} ~  \omega \non
&&=24 (1 - 136 z + 18480 z^2 - 2511104 z^3 + 341214464 z^4)+\cdots,
\ee
in complete agreement with (\ref{oriYukawa25}).

%%%%%%%%%%%%%%%%%%%%%

\section{Summary}
We studied genus-zero Gromov--Witten invariants of CY 3-folds
defined as complete intersections in Grassmannians
by using equivariant A-twisted GLSM on $S^2$. 
The Yukawa coupling
can be calculated
from the cubic correlation function of the scalar in the vector multiplet.  
In order to obtain the Yukawa coupling in the flat coordinate,
we have to compute the mirror map,
which gives the complex structure moduli
as a function of the flat coordinate.
The mirror map can be computed
from the $Z$-function
appearing in the factorization of correlation functions.
We have also studied Seiberg-like duality
between GLSMs with different ranks.
We studied only the cases when the gauge group is $U(n)$,
and it would be interesting to extend our analysis
to other gauge groups and quiver gauge theories.

Cohomological Yang--Mills theories on curved backgrounds
have recently been studied
by coupling to background topological gravity in \cite{Bae:2015eoa},
which includes   
supersymmetric background studied in \cite{Closset:2014pda}.  
It is also interesting to perform the supersymmetric localization computation
for GLSMs on these backgrounds, and
figure out their interpretation as low energy target space geometry.

%%%%%%%%%%%%%%%%%%%%%%%%%%%%%%%%%%%%%%%%%
%%%%%%%%%%%%%%%%%%%%%%%%%%%%%%%%%%%%%%%%

%%%%%%%%%%%%%%%%%%%%%%%%%%%%%%%%%%%%%%%%%%%%%%%%%%%%
%%%%%%%%%%%%%%%%%%%%%%%%%%%%%%%%%%%%%%%%%%%%%%%%%%%%
%%%%%%%%%%%%%%%%%%%%%%%%%%%%%%%%%%%%%%%%%%%%%%%%%%%%%

\subsection*{Acknowledgment}
We thank Daisuke~Inoue,
Atsushi~Ito and
Makoto~Miura
for sending a table of CY 3-folds in Grassmannians
and their Yukawa couplings.
We also thank Bumsig~Kim, Dario~Rosa, Antonio~Sciarappa and  Katsuyuki~Sugiyama for valuable discussions and comments.

%%%%%%%%%%%%%%%%%%%%%%%
%%%%%%%%%%%%%%%%%%%%%%%
%%%%%%%%%%%%%%%%%%%%%%%
%%%%%%%%%%%%%%%%%%%%%%%%%%%%
%\newpage

\end{document}